\documentclass[aps, prd, 10pt, twocolumn, superscriptaddress,noshowpacs, preprintnumbers, longbibliography,nofootinbib,bibnotes,hyperref,floatfix]{revtex4-2}
\usepackage{tikz-feynman}
\usepackage{tikz}
\usepackage{amsmath}
\usepackage{amsfonts}
\usepackage{amssymb}
\usepackage{bbold}
\usepackage{epsfig}
\usepackage{graphicx}
\usepackage{bm}
\usepackage{array}
\usepackage{hyperref}
\usepackage{listings}
\usepackage{color}
\usepackage{float}
\usepackage{booktabs}
\usepackage{url} 
\usepackage[normalem]{ulem}
\usepackage[dvipsnames]{xcolor}

\usepackage{orcidlink}

\begin{document}
\title{Steady state of fast-oscillating neutrinos  in an inhomogeneous medium}

\author{Manuel Goimil-Garc\'ia \orcidlink{0009-0001-0518-9274}}
 \affiliation{Niels Bohr International Academy \& DARK, Niels Bohr Institute,\\University of Copenhagen, Blegdamsvej 17, 2100 Copenhagen, Denmark}

\author{Irene Tamborra \orcidlink{0000-0001-7449-104X}}
\affiliation{Niels Bohr International Academy \& DARK, Niels Bohr Institute,\\University of Copenhagen, Blegdamsvej 17, 2100 Copenhagen, Denmark}

\date{\today}

\begin{abstract}
The streaming of neutrinos in  an inhomogeneous medium is known to affect the physics of flavor conversion. We employ an ensemble of single-crossed angular distributions and explore the physics of flavor conversion, while neutrinos propagate across a one-dimensional inhomogeneous medium. The  advective term in the neutrino equations of motion is responsible for the  cascade of flavor waves  towards ever smaller spatial and angular scales. However, as the system evolves, perturbations with large wavenumbers are damped, with a resulting smearing of the flavor configuration. We provide a simple recipe that allows to forecast the steady-state flavor configuration achieved by  neutrinos  without solving their kinetic equations. In particular, we find that flavor equipartition on one side of the angular spectrum and the cancellation of the spectral crossing in the lepton number distributions,  proposed in the literature as generic flavor outcome, is a special solution depending on the degree of neutrino-antineutrino asymmetry. This work constitutes a step forward towards the development of semi-analytic schemes to account for flavor conversion physics in hydrodynamic simulations of core-collapse supernovae and neutron-star merger remnants.  
\end{abstract}

\maketitle

\section{Introduction}
The neutrino density is extremely large in core-collapse supernovae and neutron-star merger remnants; despite the weakness of their interactions,  neutrinos play an essential role in  the supernova explosion mechanism as well as in the nucleosynthesis~\cite{Janka:2025tvf,Burrows:2020qrp,Tamborra:2024fcd,Vitagliano:2019yzm,Raffelt:2025wty,Fischer:2023ebq}. In such neutrino-dense media,  the evolution of flavor depends on  the refraction of neutrinos on  electrons and other neutrinos  in the background medium~\cite{Duan:2010bg,Mirizzi:2015eza,Tamborra:2020cul,Johns:2025mlm,Volpe:2023met}. 

In the source core, as neutrinos decouple, their flavor-dependent angular distributions develop crossings in the electron-muon lepton number (ELN-XLN); this condition may be responsible for inducing flavor conversion, even in the limit of vanishing vacuum frequency (fast flavor conversion)~\cite{Izaguirre:2016gsx,Morinaga:2021vmc,Padilla-Gay:2021haz,Fiorillo:2023hlk,Fiorillo:2024dik}. Various methods have been adopted to identify ELN-XLN crossings in hydrodynamic simulations, aimed at understanding where favorable conditions exist for fast flavor instabilities~\cite{Shalgar:2019kzy,DelfanAzari:2019tez,Akaho:2022zdz,Akaho:2023brj,Glas:2019ijo,Nagakura:2021hyb,Akaho:2023brj,Morinaga:2019wsv,Abbar:2020qpi,Capozzi:2020syn,Abbar:2023zkm,Cornelius:2025tyt,Cornelius:2025nvd}. However, despite the presence of ELN-XLN crossings, minimal flavor conversion may occur according to the shape of the angular distributions~\cite{Padilla-Gay:2021haz}.  Flavor conversion in the core of neutrino-dense sources can also be seeded by slow neutrino self-interaction, beyond-mean-field effects, or collisional instabilities~\cite{DedinNeto:2023ykt,Fiorillo:2025gkw,Goimil-Garcia:2024wgw, Johns:2021qby}.

  The seven-dimensional  solution of the neutrino equations of motion is hindered by the large gradients characterizing the quantities entering the equations.
Therefore,  state-of-the-art solutions of the neutrino kinetic equations  invoke symmetry assumptions  and   
attempt to solve the neutrino kinetic equations in small spatial domains~\cite{Wu:2021uvt,Richers:2021nbx,Richers:2021xtf,Zaizen:2021wwl,Xiong:2024tac,Shalgar:2022lvv,Shalgar:2022rjj,Cornelius:2023eop,Shalgar:2024gjt,Cornelius:2024zsb,Shalgar:2025oht,Nagakura:2022kic,Nagakura:2022qko}.
Understanding and  forecasting the quasi-steady state flavor configuration achieved by neutrinos is crucial  to incorporate flavor conversion physics in hydrodynamical simulations of supernovae and neutron-star merger remnants. This is especially relevant since a growing body of  work   highlights the potentially major role of flavor conversion on  the  source physics and nucleosynthesis~\cite{Ehring:2023lcd,Ehring:2023abs,Nagakura:2023mhr,Wang:2025nii,Wu:2017drk,Just:2022flt,George:2020veu,Lund:2025jjo,Qiu:2025kgy}. 

For an azimuthally symmetric and homogeneous (anti)neutrino ensemble, a simple criterion to predict the amount of flavor conversion  was proposed relying on the analogy of the neutrino kinetic equations with those of a gyroscopic pendulum~\cite{Padilla-Gay:2021haz}. In this case, the pendulum spin is provided by the real part of the eigenfrequency computed by the normal-mode analysis and allows us to predict the overall amount of flavor conversion. Alternative  approaches to forecast the asymptotic flavor configuration have been proposed relying on the Bhatnagar-Gross-Krook relaxation time anzatz~\cite{Liu:2025tnf},  
 exploiting the fact that ELN-XLN crossings vanish in the aftermath of flavor conversion~\cite{Zaizen:2022cik,Zaizen:2023ihz,Xiong:2021dex,Xiong:2024pue}, or employing neural networks~\cite{Abbar:2023ltx}. In parallel, subgrid methods are under development to facilitate the implementation of such simple schemes in hydrodynamic simulations of supernovae and neutron-star merger remnants~\cite{Johns:2024dbe,Nagakura:2023jfi,Nagakura:2022xwe,Johns:2025yxa}.

In this paper, we focus on an azimuthally symmetric, inhomogeneous (anti)neutrino ensemble and investigate the role of neutrino propagation in determining the quasi-steady state flavor configuration in the presence of  flavor conversion. The streaming of neutrinos favors the spatial spread of  regions affected by flavor conversion~\cite{Cornelius:2023eop,Shalgar:2019qwg,Padilla-Gay:2020uxa}. Moreover,  flavor waves cascade down to ever smaller  scales,  breaking the formal analogy with a gyroscopic pendulum. Relying on a suite of  ELN-XLN flavor configurations, we show that the ELN-XLN crossings are erased as a result of flavor conversion only for a subset of ELN-XLN configurations, and flavor equipartition on one side of the ELN-XLN angular spectrum is not a general outcome, as otherwise reported in the literature. Building on these results, we outline a general method to account for flavor conversion in an inhomogeneous neutrino ensemble without solving the kinetic equations.

This paper is organized as follows. In Sec.~\ref{sec:eom}, we introduce the neutrino equations of motion and investigate the role of the advective term in such equations, as well as its interplay with fast flavor instabilities. Our system setup is outlined in Sec.~\ref{sec:setup}. Then, we present our findings on the impact of advection on  flavor evolution in the linear and non-linear regimes in Sec.~\ref{sec:flavor_ev}. We forecast the steady-state flavor configuration and explore the dependence of the latter on the features of the flavor-dependent angular distributions in Sec.~\ref{sec:forecast}. We discuss our findings 
in Sec.~\ref{sec:discussion}, before  
concluding in Sec.~\ref{sec:conclusions}. Finally, we explore the impact of the initial perturbation on the steady-state flavor configuration in Appendix~\ref{app:a}.

\section{Neutrino equations of motion}
\label{sec:eom}
In this section, we introduce the neutrino kinetic equations. We then investigate the role of the advective term in the equations of motion.
\subsection{Neutrino kinetic equations}
The neutrino flavor content is described using the density matrix, $\rho_{\mathbf{p}}(\mathbf{r},t)$; the latter  depends on the neutrino momentum $\mathbf{p}$, the spatial coordinate $\mathbf{r}$, and time $t$. Similarly, the antineutrino density matrix is $\bar{\rho}_{\mathbf{p}}(\mathbf{r},t)$. The diagonal elements ($\rho_{\alpha\alpha}$, with $\alpha = e$ or $x$ in the two-flavor approximation) of the density matrix represent the distribution functions of neutrinos with flavor $\nu_\alpha$. The off-diagonal elements, $\rho_{\alpha\beta}$, encode the coherence between the neutrino flavors $\nu_\alpha$ and $\nu_\beta$.

The evolution of the density matrix is given by the  kinetic equations
\begin{equation}
\label{eq:QKE}
    i(\partial_t+\mathbf{v} \cdot \partial_\mathbf{ r})\rho_\mathbf{ p}(t,\mathbf{r})  = [H_\mathbf{{p}}(t,\mathbf{r}), \rho_\mathbf{p}(t,\mathbf{r})]\, ,
\end{equation}
with the neutrino velocity being $\mathbf{v}=\mathbf{p}/|\mathbf{p}|$. An analogous equation holds for antineutrinos. In the above equation, non-forward neutrino collisions with the background medium have been neglected, although they could be responsible for affecting flavor conversion or triggering new flavor instabilities~\cite{Hansen:2022xza,Johns:2021qby,Padilla-Gay:2022wck}. The left-hand side of Eq.~\eqref{eq:QKE} describes the evolution of the local neutrino field as a function of time, including the advection of particle density and flavor coherence. The right-hand side of Eq.~\eqref{eq:QKE} changes the flavor content of the ensemble via coherent interactions, while preserving the total number of particles. Hereafter, we neglect vacuum oscillations and the matter effect, focusing  on coherent neutrino-neutrino forward scattering. The corresponding Hamiltonian is:
\begin{equation}
{H}_{\mathbf{p}}(t,{\mathbf{r}}) = \mu \int \frac{d^3\mathbf{q}}{(2 \pi)^3} [\rho_{\mathbf q}(t,\mathbf{r})-\bar\rho_{\mathbf q}(t,\mathbf{r})] (1-\mathbf{v}_{\mathbf{q}}\cdot\mathbf{v}_\mathbf{p})\, ,
\end{equation}
where $\mu \equiv \sqrt{2} G_F n_{\nu_e}$ is the self-interaction coupling strength, with $n_{\nu_e}$ being the electron-neutrino density.

For simplicity, we assume mono-energetic neutrinos  and that our system is axisymmetric. Therefore, the only coordinates entering Eq.~\eqref{eq:QKE} are the position ($r$) and the velocity along the symmetry axis ($v = \cos\theta$, with $\theta$ being  the propagation angle). Flavor evolution can be  described by the  polarization vector $\mathbf{D}_v$:
\begin{equation}
\label{eq:polarization-def}
    D_v =\frac{1}{2}[\mathrm{Tr}(D_v)+\mathbf{D}_v\cdot\boldsymbol{\sigma}]\, ,
\end{equation}
where $D_v=\rho_v-\bar{\rho}_v$ is the lepton-number matrix for (anti)neutrinos with velocity $v$, and $\boldsymbol{\sigma}=(\sigma_x,\sigma_y,\sigma_z)$ are the Pauli matrices. 

The vector $\mathbf{D}_v$ can be expanded in multipoles:  
\begin{equation}
    \mathbf{D}_v=\sum_{m=0}^{\infty}\left(m+\frac{1}{2}\right)L_m(v)\mathbf{D}_m\, ,
\end{equation}
where $L_n(v)$ is the Legendre polynomial of order $n$, normalized so that $\int \mathrm{d}v\,L_n(v)L_m=2/(2m+1)$. Hence, Eq.~\eqref{eq:QKE} leads to: 
\begin{subequations}
\label{eq:multipole}
\begin{alignat}{2}
    &\partial_t\mathbf{D}_v +v\partial_r\mathbf{D}_v = \mu  (\mathbf{D}_0-v\mathbf{D}_1)\times \mathbf{D}_v\, ,\label{eq:polarization-ev}\, \\
    &\partial_t\mathbf{D}_l+\partial_r\mathbf{T}_l =\mu (\mathbf{D}_0\times \mathbf{D}_l-\mathbf{D}_1\times \mathbf{T}_l)\, ,\label{eq:multipole-ev} 
    \end{alignat}
\end{subequations}
where  $\mathbf{T}_l \equiv {l}/({2 l + 1}) \mathbf{D}_{l-1}+({l+1})/({2 l + 1})\mathbf{D}_{l+1}$, and we  use the following identity to go from Eq.~\eqref{eq:polarization-ev} to \eqref{eq:multipole-ev}: 
\begin{equation}
\begin{split}
    \int \mathrm{d}v\,vL_m(v)L_n(v) &=\frac{2(m+1)}{(2m+1)(2m+3)}\delta_{m+1,n}\\&+\frac{2m}{(2m-1)(2m+1)}\delta_{m-1,n}\, .
    \end{split}
\end{equation}
Equation~\eqref{eq:multipole-ev} shows that the advective term couples $\mathbf{D}_l$ with $\mathbf{D}_{l\pm 1}$. In addition, a similar coupling among  multipoles is also given by the  self-interaction term~\cite{Johns:2020qsk,Raffelt:2007yz,Padilla-Gay:2021haz}. 

\subsection{Neutrino streaming}
 In a homogeneous and azimuthally symmetric medium, 
one can show that the dynamics of the neutrino ensemble is  equivalent to the one of a gyroscopic pendulum in flavor space, where $\mathbf{D}_0$ plays the role of gravity and $\mathbf{D}_1$ is the position of the mass~\cite{Padilla-Gay:2021haz, Fiorillo:2023mze} (see also Ref.~\cite{Johns:2019izj}). Hereafter, we explore how this analogy is modified once the impact of inhomogeneities on flavor evolution is accounted for.

The (anti)neutrino angular distribution is sensitive to the spatial structure of the ensemble, because the directional derivative in Eq.~\eqref{eq:polarization-ev} couples $\mathbf{D}_l$ and $\mathbf{D}_{l\pm 1}$. This coupling violates the conservation of, e.g., $\mathbf{D}_0(t,r)$ and $|\mathbf{D}_1(t,r)|$, that  holds when the spatial gradient is negligible~\cite{Padilla-Gay:2021haz}. Hence, inhomogeneous media cannot sustain pendulum-like behavior. Due to advection, polarization vectors in different spatial locations become coupled 
and collective flavor conversion cascades towards ever smaller angular and spatial scales. 

For example, we  consider a  box of size $L$ that is subject to local perturbations. Expressing each polarization vector  as $\mathbf{D}_v(t,r)=\sum_m\mathbf{\hat{D}}_v(t,k_m)\exp(ik_mr)$, where $k_m=2\pi m/L$, and  inserting this definition in Eq.~\eqref{eq:polarization-ev}, we obtain: 
\begin{eqnarray}
        &{\mu^{-1}}& (\partial_t-vk_m)\mathbf{\hat{D}}_v(k_m) =(\langle \mathbf{D}_0\rangle-v\langle \mathbf{D}_1\rangle)\times \mathbf{\hat{D}}_v(k_m)\label{eq:fourier-polarization-ev}\\
        &+&\sum_{j\neq 0}(\mathbf{\hat{D}}_0-v\mathbf{\hat{D}}_1)(k_j)\times \mathbf{\hat{D}}_v(k_m-k_j)\, ,\nonumber
\end{eqnarray}
 where $\langle \mathbf{D}_l\rangle \equiv \mathbf{\hat{D}}_n(k_m=0)$ denotes  spatially averaged quantities across the box. The second term on the right-hand side of Eq.~\eqref{eq:fourier-polarization-ev} couples all Fourier modes and changes both  direction and  module of each $\mathbf{\hat{D}}_v(k_m)$. Hence, from here, we  see that  coherent interactions can stimulate the growth of short-length inhomogeneities.

\section{System setup}
\label{sec:setup}
In order to explore the interplay between inhomogeneities, the shape of the  angular distributions, and neutrino self-interaction, we consider the following suite of (anti)neutrino angular distributions at $t=0$: 
\begin{subequations}
\label{eq:ini_conditions}
    \begin{alignat}{2}
        \rho_{v,ee}(0,r) & = 0.5\, ,\\
        \bar{\rho}_{v,ee}(0,r) &= 0.1 +\frac{\alpha}{N}\,\exp\left( -\frac{(v-1)^2}{2\sigma^2}\right)\, ,\\
                \rho_{v,xx}(0,r)&= \bar{\rho}_{v,xx}(0,r) = 0\, ,
    \end{alignat}
\end{subequations}
where $N=\int_{-1}^{+1}\mathrm{d}v\,\exp\left( -{(v-1)^2}/{2\sigma^2}\right)$. These matrices are related via Eq.~\eqref{eq:polarization-def} to  $\mathbf{D}_v$.  They describe homogeneous ELN-XLN distributions ($D_v^z$), with $\nu_e$ assumed to be isotropic  and $\bar{\nu}_e$ preferentially traveling  towards positive $r$ ($v>0$). The width of the forward peak of $\bar{\rho}_{v,ee}$ is determined by $\sigma$, while $\alpha$ governs the average ELN-XLN: 
\begin{equation}
\label{eq:D0z}
\langle D_0^z\rangle  = \frac{1}{L}\int_0^L\mathrm{d}r\,\int_{-1}^{+1} \mathrm{d}v\, {D^z_v(0,r)} =0.8-\alpha\, ,
\end{equation}
which is a conserved quantity, according to Eq.~\eqref{eq:fourier-polarization-ev}. 
In order to explore different ELN-XLN configurations, we let $\alpha$ and $\sigma$ vary between $[0.4,0.7]$ and $[0.45,0.7]$, respectively. All  ELN-XLN distributions in Eq.~\eqref{eq:ini_conditions} have one angular crossing; i.e., each ELN-XLN angular distribution has  a neutrino velocity $v_c$ such that $D^z_{v<v_c}>0$ and $D^z_{v\geq v_c}\leq 0$. 

We solve Eq.~\eqref{eq:QKE} for neutrinos and antineutrinos in a box of width $L$ with periodic boundary conditions [i.e., $\rho_v(t,0)=\rho_v(t,L)$]. We consider  inhomogeneous perturbations in the off-diagonal terms of the density matrices to trigger flavor conversion: 
\begin{equation}
\label{eq:perturbation}
    \rho_{v,xe}(0,r)=A\exp\left(-\frac{(r-L/2)^2}{2s_p^2}\right)\, .
\end{equation}
In all our examples, we use $A = 10^{-5}$,  $s_p = 9 \mu^{-1}$, $L=900 \mu^{-1}$, and compute the evolution of the density matrices up to $t_f = 900 \mu^{-1}$. This parameter choice  allows to  follow the evolution of the spatial perturbation and flavor evolution across the box, while preventing inhomogeneities to spread at unmanageably small scales. 

We discretize the angular coordinate into $100$ bins and ensure that the spatial resolution of our simulations is enough to resolve the fastest-growing eigenmode in our parameter space~\cite{Nagakura:2025brr}; cf.~also Sec.~\ref{sec:lsa}. As summarized in Table~\ref{tab:1}, we select four benchmark sets of angular distributions, C1--C4, to describe the phenomenology of flavor conversion in the presence of neutrino streaming.

\begin{table}[H]
    \centering
    \caption{Parameters of our benchmark angular distributions; see Eq.~\eqref{eq:ini_conditions} and the right panel of Fig.~\ref{fig:1}.}
   \begin{tabular}{c|cc}
    \toprule 
    Distribution & $\alpha$ & $\sigma$ \\ \midrule
       C1  & 0.50 & 0.50\\
       C2  & 0.50 & 0.65\\
       C3  & 0.55 & 0.55\\
       C4  & 0.60 & 0.60 \\ \bottomrule
    \end{tabular}
    \label{tab:1}
\end{table}

\section{Flavor conversion in the presence of advection}
\label{sec:flavor_ev}
In this section, we explore  the response of the neutrino ensemble to inhomogeneous perturbations.
First,  we  investigate the linear regime  following a semi-analytical approach, then  we explore numerical solutions of the (anti)neutrino equations of motion.

\subsection{Linear regime}
\label{sec:lsa}
The stability of the neutrino system against perturbations of characteristic length $k_n$ can be diagnosed based on the shape of the angular ELN-XLN  distribution~\cite{Izaguirre:2016gsx, Yi:2019hrp, Fiorillo:2024dik}. We insert $\hat{D}_v^z(k_n)=\langle {D}_v^z\rangle \delta_{n0}$ in Eq.~\eqref{eq:fourier-polarization-ev} and look for solutions of the form $[\hat{D}_{v}^x(t,k_n)+i\hat{D}_v^y(t,k_n)]\propto \exp[i(k_nr-\omega t)]$--hereafter the symbol $\langle ...\rangle$ is used to denote the spatial average across the box. 

The eigenfrequency $\omega(k_n) =\omega_R(k_n)+i\omega_I(k_n)$ is a solution of 
\begin{equation}
    \label{eq:dispersion relation}
    (I_0-1)(I_2+1)-I_1^2=0\, ,
    \end{equation}
    where
    \begin{equation}
    I_n(k_m) = \int \mathrm{d}v\,\frac{v^n\langle D_v^z\rangle }{\omega+\mu \langle D_0^z\rangle -v(k_m+\langle D_1^z\rangle)}\, .
    \end{equation}
    If $\omega_I >0$,  the (anti)neutrino ensemble  undergoes a fast flavor instability.
    Equation~\eqref{eq:dispersion relation} 
allows for such runaway modes if there is an angular crossing in the ELN-XLN distribution~\cite{Morinaga:2021vmc}. 
 \begin{figure*}
\centering
\includegraphics[width=\linewidth]{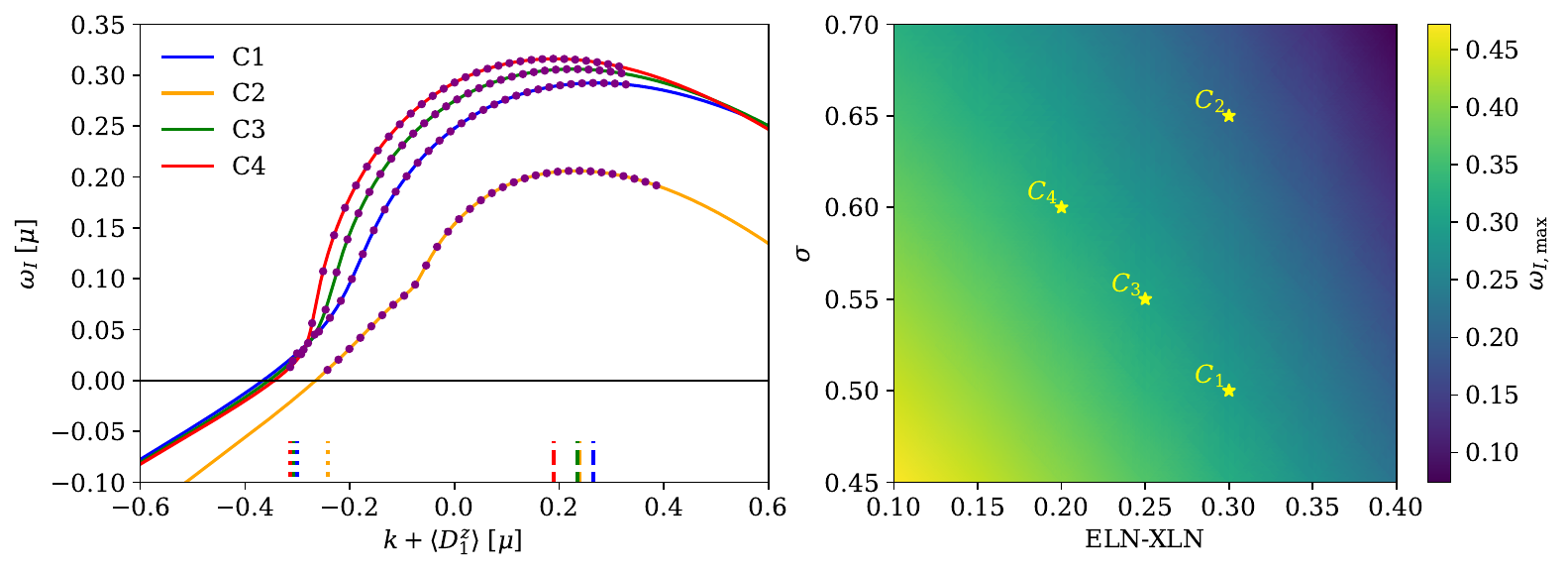}
\caption{\textit{Left:} Growth rate of flavor instabilities as a function of their effective wavenumber, $K=k+\langle D^z_1\rangle$, where $\langle D_1^z\rangle$ is the space-averaged ELN-XLN flux, for the angular distributions C1--C4 defined by Eqs.~\eqref{eq:ini_conditions} and Table~\ref{tab:1}. The vertical dotted (dashed) lines signal the location of the homogeneous (fastest-growing) mode. The dots on each $\omega_I(k)$ curve mark the Fourier modes $k_m=2\pi m/L$ for $0\leq m\leq 30$; this  highlights that periodic boundary conditions impose discrete spectra on the solutions of the equations of motion. \textit{Right:} Maximum growth rate of the flavor instability for the single-crossed angular distributions defined by Eqs.~\eqref{eq:ini_conditions}, in the plane spanned by the space-averaged ELN-XLN ($\langle D_0^z\rangle \propto \alpha$, cf.~Eq.~\eqref{eq:D0z}) and the width of the $\bar{\nu}_e$ distribution ($\sigma$). For all cases in this parameter space, the fastest-growing mode is inhomogeneous ($k\neq 0$).}
\label{fig:1}
\end{figure*}

The left panel of Fig.~\ref{fig:1} shows the growth rate  of the flavor instability
as a function of the effective wavenumber of the perturbation, $K\equiv k+\langle D_1^z\rangle$,  for the  ELN-XLN configurations C1--C4 summarized in Table~\ref{tab:1}. The behavior of these four (anti)neutrino configurations in the linear regime is dominated by inhomogeneous modes: $\omega_I(K)$ has a maximum at $K_\mathrm{max}\neq \langle D_1^z\rangle$, which  is an order of magnitude higher than the homogeneous growth rate (cf.~dotted vs.~dashed vertical lines in Fig.~\ref{fig:1}). Deeper angular crossings (smaller $\sigma$) correlate with higher $\omega_I$ across all wavenumbers and  a milder dependence on $K$. A smaller  ELN-XLN (higher $\alpha$)  leads to a higher $\omega_I$ when $K\neq \langle D_1^z\rangle$. This trend is generalized  in the right panel of Fig.~\ref{fig:1}, which shows the maximum growth rate for each flavor configuration in our suite of angular distributions (cf.~Sec.~\ref{sec:setup} and Eq.~\eqref{eq:ini_conditions}).

The stability conditions of the flavor pendulum are easier to break in very asymmetric ensembles: the homogeneous mode is  unstable when the ELN-XLN crossing is prominent enough, but not so deep as to have comparable angle-integrated lepton numbers for $v<v_c$ and $v>v_c$. Consequently, it is possible to find ELN-XLN configurations such that the flavor pendulum would be stable, if the system were homogeneous, but become unstable if neutrino streaming is taken into account. Our parameter space excludes such cases, but our choice of ELN-XLN angular distributions  highlights that the interplay between advection and neutrino self-interaction can magnify small perturbations with relatively large wavelengths, triggering collective flavor instabilities.

\subsection{Non-linear regime}
\label{sec:num}

\begin{figure*}
    \centering
    \includegraphics[width=1\textwidth]{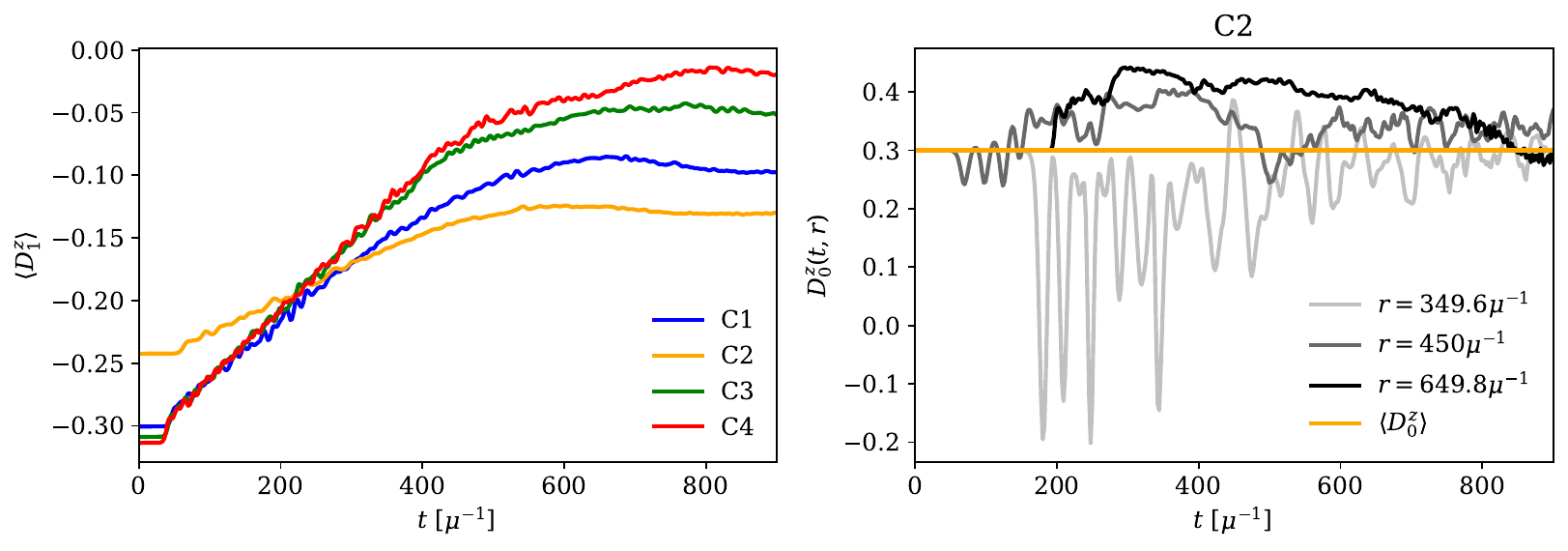}
    \caption{\textit{Left:}  Evolution of the space-averaged  lepton-number flux, $\langle D_1^z\rangle $, for the ELN-XLN distributions C1--C4 highlighted in Fig.~\ref{fig:1} (see also Table~\ref{tab:1}). The system enters the non-linear regime  at $t~\sim 100 \mu^{-1}$ for all our benchmark ELN-XLN distributions. In a homogeneous medium, the evolution would be periodic, but advection forces the polarization vectors away from the $z$-axis, breaking such periodicity. \textit{Right:} Angle-integrated ELN-XLN number, $D_0^z$, as a function of time for the ELN-XLN configuration C2. The orange line shows the space-averaged $D_0^z$, which is conserved according to Eq.~\eqref{eq:fourier-polarization-ev}; the remaining lines show the evolution of the local lepton number extracted at different spatial locations in the box ($r = 349.6 \mu^{-1}$, $450 \mu^{-1}$, and $649.8 \mu^{-1}$, from light to dark gray, respectively). Close to the peak of the initial perturbation (located at $r = 450 \mu^{-1}$),  $D_0^z$ initially shows signs of periodicity  (cf.~light gray curves); such periodicity, however, disappears at later times as inhomogeneities  cascade to small scales. Farther away from $r = 450 \mu^{-1}$ (cf.~black line), the onset of the non-linear phase is delayed  and no periodic features appear.  
    }
    \label{fig:2}
\end{figure*}

The impact of advection on flavor conversion  becomes more prominent once the system enters its  non-linear phase. 
The left panel of Fig.~\ref{fig:2} shows the evolution of  $\langle D_1^z\rangle$  for the initial angular configurations C1--C4 summarized in Table~\ref{tab:1}. The linear regime lasts for $t \lesssim 100\mu^{-1}$: in this time interval, the unstable Fourier modes linked to the perturbation in Eq.~\eqref{eq:perturbation} grow exponentially, but flavor conversion does not kick in because the transverse components $D_1^{xy}$ are still small compared to $D_1^z$. Once $D_1^{xy}$ becomes large enough, $\mathbf{D}_1$ suddenly tilts away from the $z$-axis. At this point, the system enters the non-linear regime and the convolution of the modes with $k_m\neq 0$ in Eq.~\eqref{eq:fourier-polarization-ev} becomes significant.

Early in the non-linear phase,  the evolution of the ensemble mildly depends on the shape of the initial perturbation, however the steady-state flavor configuration is negligibly affected by the choice of the perturbation, as discussed in Appendix~\ref{app:a}.  Equation~\eqref{eq:perturbation} describes a Gaussian centered around $r=L/2$. Flavor instabilities first manifest in the middle of the box, in the proximity of $L/2$, and then spread due to advection. In the vicinity of the initial perturbation, non-linear flavor conversion begins earlier than in other regions of the box. Here, the Fourier spectrum of the ensemble is still relatively narrow and dominated by the homogeneous mode, so $D_0^z$ oscillates almost periodically shortly after the onset of flavor conversion, as shown in the right panel of  Fig.~\ref{fig:2}. This pendulum-like behavior is destroyed as the coupling between length scales in Eq.~\eqref{eq:fourier-polarization-ev} continues to broaden the spectrum. For spatial locations  farther away from the peak of the initial perturbation, which are  reached by the time inhomogeneities have  cascaded down to smaller scales, the non-linear regime of flavor conversion  begins later and no periodic features are observed in $D_0^z$ (cf.~dark vs.~gray lines in the right panel of Fig.~\ref{fig:2}). 

Inhomogeneities on short length scales correspond to terms with rapidly varying phases in the Fourier transform of the polarization vectors. Hence, integrals of $\mathbf{D}_v(t,r)$ over the spatial coordinate $r$, which we refer to as coarse-grained polarization vectors,  become incoherent sums of $k$-modes. 
As an immediate consequence, space-averaged quantities  do not oscillate at the onset of the non-linear regime (cf.~left panel of Fig.~\ref{fig:2}); instead, the local polarization vectors $\mathbf{D}_v^z(t,r)$, which point in different directions and enter the non-linear regime at different times, make $\langle \mathbf{D}_{l>1}\rangle$ drift away from the $z$-axis.  As the coherence between neighboring polarization vectors decreases, so does the effective amount of flavor conversion, and $\langle D_1^z\rangle$ approximately plateau at $t \gtrsim 500\mu^{-1}$, as shown in the left panel of Fig.~\ref{fig:2}. 

 \begin{figure*}
    \centering
    \includegraphics[width=\textwidth]{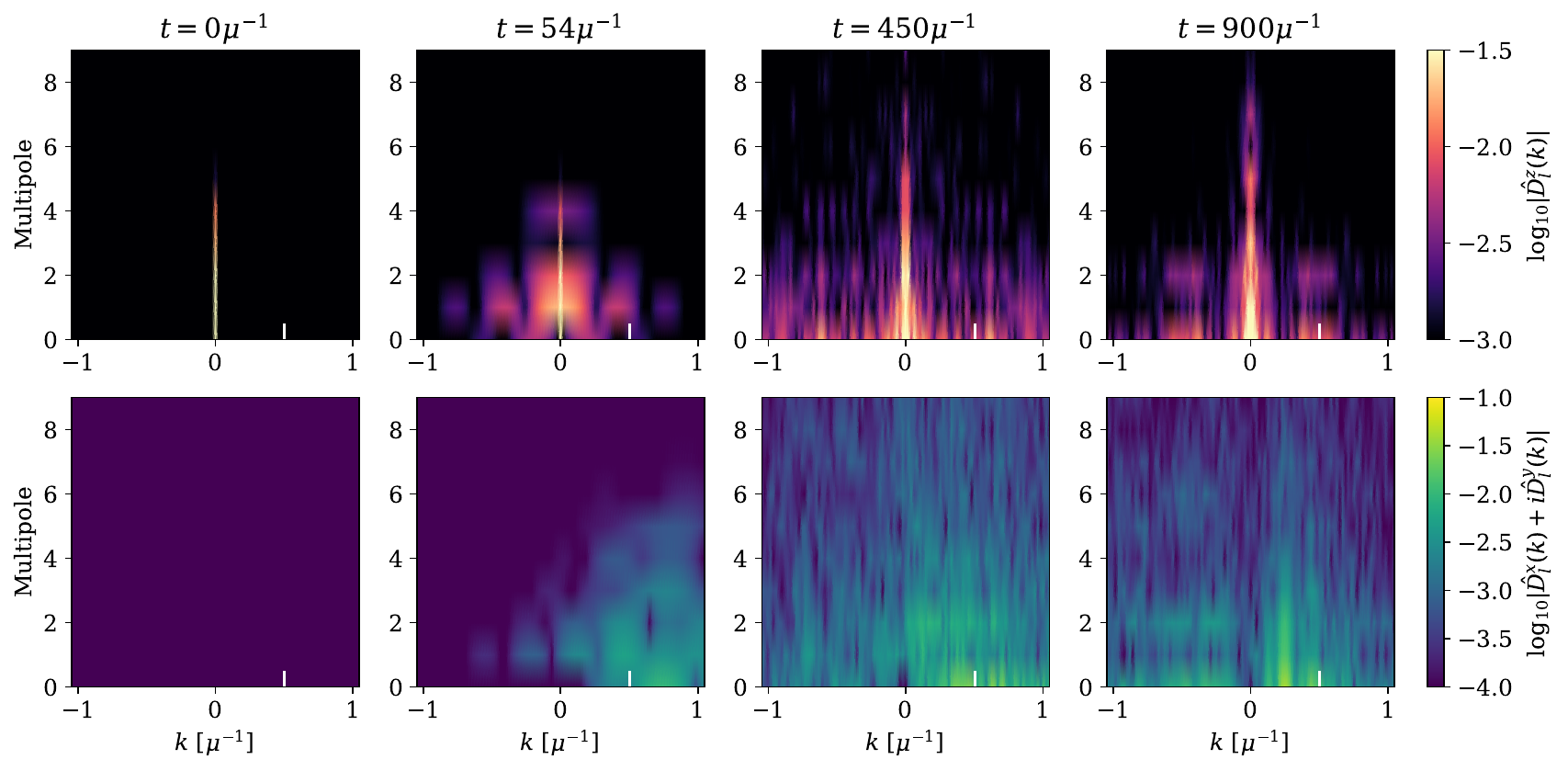}
    \caption{Snapshots of the Fourier ($x$-axis) and multipolar ($y$-axis, for $\hat{D}_l(t)$ and $0\leq l\leq 9$) decomposition for the neutrino ensemble C4 extracted at $t=0$, $54 \mu^{-1}$, $450 \mu^{-1}$, and $900 \mu^{-1}$, from left to right, respectively. The top (bottom) panels represent the (off-)diagonal components of $D_l(t,r)$. Flavor instabilities trigger a cascade towards small spatial and angular scales (cf.~middle panels), which  subsides as the ELN-XLN distribution approaches a steady state (cf.~right panel). The white line at $k\sim 0.5\mu^{-1}$ marks the position of the fastest-growing Fourier mode. }
    \label{fig:3}
\end{figure*}
 The evolution of one of our benchmark angular configurations C4 can be summarized as in Fig.~\ref{fig:3}, which shows the Fourier spectrum of the multipoles $D_l^z(t,r)$ for $0\leq l\leq 9$ on top and the corresponding off-diagonal components of $D_l(t,r)$ on the bottom. The top panels show that the ELN-XLN distribution is initially homogeneous, and the Fourier  spectrum only contains the $k=0$ mode, but all multipoles are non-zero due to the  shape of the angular distribution. Then, because of flavor conversion,  the  Fourier modes with $k\neq 0$ are populated; local fluctuations  stimulate the growth of higher-order multipoles, changing the shape of the  (anti)neutrino angular distribution. However, this cascade does not continue indefinitely. For $l\geq 4$, the inhomogeneous modes eventually start decaying due to  smearing effects of neutrino streaming; this correlates with flavor conversion being damped at $t \gtrsim 500\mu^{-1}$ (see also the left panel of Fig.~\ref{fig:2}). Notably, the amplitude of the fastest-growing mode due to the inhomogeneity is several orders of magnitude smaller than that of the homogeneous mode at the end of the simulation. As visible from the top right panel of Fig.~\ref{fig:3}, the final state of the neutrino ensemble, which we discuss in Sec.~\ref{sec:forecast}, is dominated by modes with low wavenumber. On the other hand, the bottom panels of Fig.~\ref{fig:3} show that $D_l^x(t,r) + i D_l^y(t,r)$ continues to fluctuate across angular and spatial scales, with the  homogeneous mode being subdominant.

\section{Forecast of the steady-state flavor configuration}
\label{sec:forecast}
In this section, we provide a criterion to forecast the steady state~\footnote{Note that, because of small fluctuations affecting the asymptotic flavor configuration (see, e.g., Fig.~\ref{fig:3}), it is appropriate to look for the quasi-steady state achieved by the neutrino ensemble. However, we are interested in constraining  quantities averaged over the spatial domain; so small-scale fluctuations  are smeared out by the averaging process. For simplicity, we name this  the ``steady-state'' configuration of the system.} of flavor conversion for an inhomogeneous system. To this purpose, we introduce the  $\nu_e$ survival probability ($P_{ee}$) averaged over the spatial domain once the steady state has been achieved at time $t_f$:
\begin{equation}
\label{eq:Pee}
    P_{ee}(t_f, r) = \frac{\int \mathrm{d}v\, \rho_{v,ee}(t_f,r)}{\int \mathrm{d}v\,\rho_{v,ee}(0,r)}\, .
\end{equation}

The steady-state flavor configuration can be forecast by employing  the following 
empirical prescription: 
\begin{eqnarray}
\label{eq:finalstate}
&&\langle {P}_{ee}(t_f)\rangle_{\rm{steady}} =\\
&&\int \mathrm{d}v\,\left[\frac{1}{3}\Theta [-\langle D_v^z(0)\rangle ]+\left(1-\frac{2|I_-|}{3I_+} \right)\Theta[\langle D_v^z(0)\rangle]\right]\, , \nonumber
\end{eqnarray}
where $ I_\pm = \int_{-1}^{+1} \mathrm{d}v\,\Theta (\pm \langle D_v^z(0)\rangle)\langle D_v^z(0)\rangle$, and $\Theta (x)$ is the Heaviside theta function.

\begin{figure*}
    \centering
    \includegraphics[width=\linewidth]{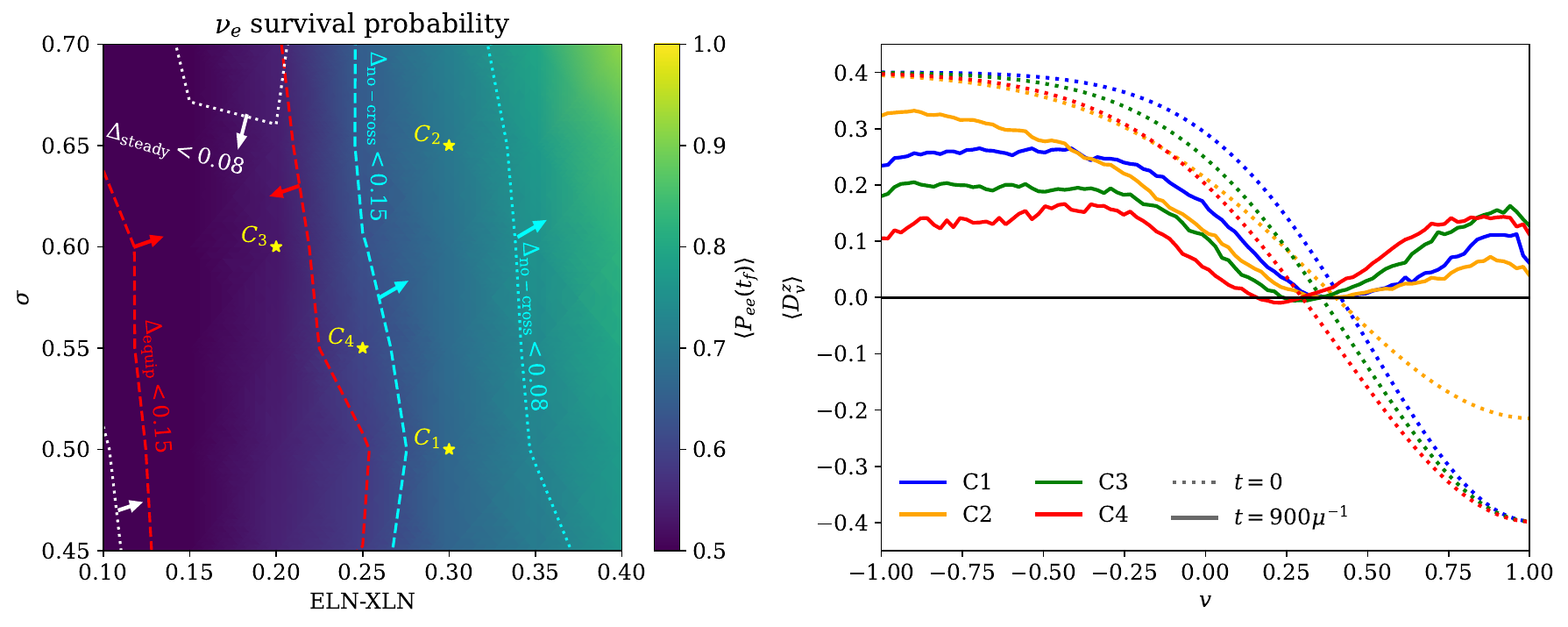}
    \caption{{\it Left:} Angle-integrated and space-averaged $\nu_e$ survival probability in the plane spanned by the average ELN-XLN ($\langle D_0^z\rangle $) and the width of the $\bar{\nu}_e$ distribution ($\sigma$).
    The white, red, and cyan contours mark the  regions with small relative error between the numerical solutions and the criteria defined by  Eqs.~\eqref{eq:finalstate}, \eqref{eq:equipartition}, and \eqref{eq:no-crossing}, respectively. As for the relative error, we use different confidence levels {(dotted contours for $\Delta \lesssim 8\%$  and dashed contours for $\Delta \lesssim 15\%$)}. Equation~\eqref{eq:finalstate} reproduces numerical results significantly better than the remaining criteria, which  model partial (and complementary) regions of our parameter space with a lower degree of accuracy. 
  Our four benchmark ELN-XLN configurations, C1--C4 (cf.~Table~\ref{tab:1}), are marked with yellow stars. 
    {\it Right:} Spatially averaged  ELN-XLN distribution for the angular configurations C1--C4. The dotted (solid) lines show the distribution at the beginning (end) of each simulation. We note that the  part of the angular distribution with $v>v_c$ is not driven to $0$, but changes its sign as a result of flavor conversion.
    \label{fig:4}}
\end{figure*}

The left panel of Fig.~\ref{fig:4} shows $\langle P_{ee}(t_f)\rangle$, obtained by averaging Eq.~\eqref{eq:Pee} over the simulation box. The white dashed contour highlights  the validity range of Eq.~\eqref{eq:finalstate}. In particular, we plot the relative error $\Delta_{\rm{steady}} =|\langle P_{ee}(t_f)\rangle-\langle P_{ee}(t_f)\rangle_{\rm{steady}} |/\langle P_{ee}(t_f)\rangle$  and highlight the region with  $\Delta_{\rm{steady}} \lesssim 8\%$. We can see that Eq.~\eqref{eq:finalstate} performs extremely well and allows us to  predict the final flavor configuration without solving the neutrino kinetic equations. 

{We compare Eq.~\eqref{eq:finalstate} with two alternative prescriptions, which have been considered in the literature to predict the flavor outcome. We find that the agreement of such prescriptions with our numerical calculations strongly depends on $\langle D_0^z\rangle$}.
\begin{paragraph}
        {\it Small ELN-XLN} ($\langle D_0^z\rangle <0.25$). 
       As an extreme scenario, it has been postulated that the neutrino ensemble approaches flavor equipartition because of fast flavor conversion, while conserving the lepton number~\cite{Sawyer:2005jk}. 
    In order for this to hold,  antineutrinos reach equipartition, while neutrinos evolve to preserve the angle-integrated ELN-XLN (see, e.g., case C3 in the left and right  panels of Fig.~\ref{fig:4}). Hence, 
    \begin{equation}
\label{eq:equipartition}
    \langle P_{ee}(t_f)\rangle_{\rm{equip}} =\int \mathrm{d}v\, [\langle \rho_{v,ee}(0)\rangle - \frac{2}{3}\langle \bar{\rho}_{v,ee}(0)\rangle ]\,  .
\end{equation}
The validity of this criterion is represented by the red dashed lines in the left panels of Fig.~\ref{fig:4}, where  we plot $\Delta_{\rm{equip}} \lesssim 15\%$ (for a relative error $\lesssim 8\%$, as we show for $\Delta_{\rm{steady}}$, the red contours  fall out of our parameter space; therefore this approximation would not be valid for any of the ELN-XLN distributions in our suite).
Figure~\ref{fig:4} clearly shows that flavor equipartition can only be achieved in the region  of the parameter space where $\langle D_0^z\rangle <0.25$; hence, this is not a general outcome of flavor conversion.
\end{paragraph}

\begin{paragraph}{\it Large ELN-XLN} ($\langle D_0^z\rangle >0.25$). Fast flavor conversion tends to erase the initial ELN-XLN crossing, i.e., the angular distribution tends to $D_{v}^z=0$ for $v>v_c$ and the $v<v_c$ part of the angular spectrum  adjusts accordingly to preserve the total ELN-XLN. We model the steady-state ELN-XLN distribution using a step function inspired by Refs.~\cite{Zaizen:2022cik, Zaizen:2023ihz}:
\begin{eqnarray}
\label{eq:no-crossing}
&& {\langle P_{ee}(t_f)\rangle_{\rm{no-cross}}} =\\ &&\int \mathrm{d}v\,\left[\frac{1}{2}\Theta [-\langle D_v^z(0)\rangle ] 
    +\left(1-\frac{|I_-|}{2I_+} \right)\Theta[\langle D_v^z(0)\rangle]\right] \, . \nonumber
\end{eqnarray}
This criterion corresponds to the dashed cyan contour in the left panel of Fig.~\ref{fig:4} ({the dotted contour refers to a relative error $\lesssim 8\%$}) and holds for our benchmark cases C1 and C2; see also the right panel of Fig.~\ref{fig:4}. We find that Eq.~\eqref{eq:no-crossing} only holds for $\langle D_0^z\rangle >0.25$ for our ensemble of neutrino angular distributions. Therefore, equipartition on one side of the angular crossing is not a generic outcome of fast flavor conversion, unlike suggested in Refs.~\cite{Wu:2021uvt, Richers:2022bkd, Zaizen:2022cik,Zaizen:2023ihz, Xiong:2023vcm}.
\end{paragraph}

It is important to note that Eqs.~\eqref{eq:equipartition} and \eqref{eq:no-crossing}  span opposite regions of our parameter space, as shown in the left panel of Fig.~\ref{fig:4}. A priori none of them can reliably predict the final flavor configuration for any ELN-XLN distribution. This is because flavor  equipartition tends to be achieved when the numbers of neutrinos and antineutrinos are comparable and the angular crossing is  deep. In  this case, most of the (anti)particles undergo flavor conversion. On the other hand, 
the ELN-XLN crossing is erased for large lepton numbers  (or high neutrino-antineutrino asymmetry)--a configuration which naturally leads to shallow crossings. Importantly,  the  part of the angular distribution with $v>v_c$ is not driven to $0$, but can change sign (cf.~Fig.~\ref{fig:4}, right panel). 

\section{Discussion}
\label{sec:discussion}

Equation~\eqref{eq:finalstate} offers a generic recipe  to predict the steady-state flavor configuration of an inhomogeneous system. We now put our results in the context of previous efforts to predict the asymptotic outcome of fast instabilities.

By comparing cases C1--C4 in Fig.~\ref{fig:4}, we can see that the steady-state flavor configuration depends on the ELN-XLN number ($\langle D_0^z\rangle$) and the slope of the ELN-XLN in the proximity of $v_c$. {In particular,  Eq.~\eqref{eq:finalstate} accounts for the fact that  the areas of the angle-integrated ELN-XLN distribution for $v< v_c$ and $v> v_c$ are responsible for determining the steady-state configuration based on conservation laws.} Approximating the shape of the ELN-XLN distribution may require a large number of multipoles $\mathbf{D}_l$ even in the absence of flavor conversion~\cite{Cornelius:2025tyt,Johns:2021taz};   collective instabilities in inhomogeneous media should be expected to generate non-trivial contributions at higher orders, according to Eq.~\eqref{eq:multipole-ev}.  
In this sense, attempts to  find the steady state of fast flavor conversion by truncating the multipole expansion of the density matrices  at an arbitrarily low $l$ cannot  be  predictive. 
For example, Ref.~\cite{Bhattacharyya:2022eed} considers  $\mathbf{D}_l=0$ for  $l>3$; we have tested that their prescription does not qualitatively reproduce the survival probabilities shown in Fig.~\ref{fig:4}  (results not shown here). This is due to the fact that our initial ELN-XLN distributions   include non-trivial higher-order multipoles ($l>3$), which grow significantly over time in all our selected examples (cf.~Fig.~\ref{fig:3}). 
This finding is in agreement with the conclusions reached in Refs.~\cite{Johns:2019izj, Padilla-Gay:2021haz} for a homogeneous system. 
Moreover, approximate solutions of Eq.~\eqref{eq:multipole-ev}  were derived in Ref.~\cite{Bhattacharyya:2022eed} under the  assumption that the polarization vectors are parallel to the $z$-axis at late times, which is not generally true  (cf.~Fig.~\ref{fig:3}, bottom panels).

Likewise, methods that aim to forecast the steady-state configuration  relying on assumptions about the shape of the Fourier spectrum might be justified only under special circumstances. For example, Ref.~\cite{Liu:2024nku} found an approximate solution of the (anti)neutrino equations of motion by assuming that:
\begin{equation}
    \label{eq:factorization}
    \langle D_l^aD_m^b\rangle =\langle D_l^a\rangle \langle D_m^b\rangle\,,
\end{equation}
i.e., that the convolution of inhomogeneous modes on the right-hand side of Eq.~\eqref{eq:fourier-polarization-ev} is negligible. 
We have generalized the approach introduced in Ref.~\cite{Liu:2024nku}, which considers two neutrino beams colliding head-on ($v=\pm 1$), to the case of a continuous angular distribution. We find that Eq.~\eqref{eq:factorization} holds at all times for initial configurations with shallow crossings and large neutrino-antineutrino asymmetries. However, when the ELN-XLN distribution is more forward peaked, Eq.~\eqref{eq:factorization}  is only accurate at very early and very late times (results {not} shown here). This finding can be explained by Fig.~\ref{fig:3}, where it is clear that  perturbations with large $k$ tend to subside at late times due to smearing effects induced by advection{. This results reminds us of the damping of small-scale structures numerically found in Refs.~\cite{Shalgar:2019qwg,Padilla-Gay:2020uxa}; however, we note that Refs.~\cite{Shalgar:2019qwg,Padilla-Gay:2020uxa} adopted variable neutrino densities across the simulation shell in the presence of advection, allowing for the spreading of large-scale inhomogeneities that are not present in the system setup investigated here.} 

We explore the role of inhomogeneities on the flavor evolution within a box with periodic boundary conditions. The choice of periodic boundaries is justified by our goal of searching for an  analytical method that  allows to predict the flavor configuration without solving the equations of motion. However, if a system with periodic boundaries evolves over several crossing times ($L$), neutrinos with $v\neq 0$ repeatedly encounter the same large perturbation.  Such an approach automatically leads to flavor depolarization as an artifact of the choice of   boundary conditions~\cite{Cornelius:2023eop}. 
In order to avoid this issue, we investigate flavor evolution within the timeframe $t_f=L$, so (anti)neutrinos do not cross the spatial domain  more than once.

In addition to neutrino self-interaction and advection, the steady-state flavor configuration can be affected by collisions with the background medium~\cite{Hansen:2022xza,Johns:2021qby}, as well as vacuum oscillations~\cite{DedinNeto:2023ykt,Shalgar:2020xns,Padilla-Gay:2025tko,Fiorillo:2025gkw}. We stress that the main findings of this work  serve as a step forward in our understanding of the flavor configuration in the aftermath of flavor conversion, but they could  be further modified by vacuum effects and collisions; more work is needed in this direction.

\section{Conclusions}
\label{sec:conclusions}
Our understanding of how neutrinos change their flavor in the core of neutrino-dense sources is hindered by the conceptual and technical challenges characterizing the modeling of this phenomenon. In this paper, 
we  explore the interplay between coherent neutrino-neutrino interaction and neutrino streaming.  To this purpose, we employ an ensemble of single-crossed ELN-XLN distributions and explore the neutrino flavor evolution in a one-dimensional inhomogeneous box with periodic boundary conditions. 

We find that advection triggers a cascade of flavor waves towards increasingly small scales, with the fastest growth of flavor instability being driven by inhomogeneous modes in all our examples. However,  perturbations with large wavenumbers ($k$), which can grow very rapidly early on, tend to be damped at late times; this is because of the smearing effects of convection. The asymptotic ELN-XLN distribution is thus dominated by quasi-homogeneous modes. 

We offer a simple empirical formula to forecast the steady-state flavor configuration in the aftermath of flavor conversion in an inhomogeneous system. {We aim to  predict the angle-integrated survival probability $\langle P_{ee}(t_f) \rangle$ averaged  over our simulation box, since this is the quantity expected to  be of relevance for assessing the feedback of neutrino conversion on the source physics. However, we stress that our recipe for $\langle P_{ee}(t_f) \rangle$ intrinsically depends on the shape of the angular distribution   before and after the crossing. As such, it cannot be employed accurately in the context of hydrodynamic simulations which rely on moment-based neutrino transport schemes because of their intrinsic limitations in the reconstruction of the neutrino angular distributions in post-processing (cf.~Refs.~\cite{Cornelius:2025tyt,Johns:2021taz}). For} initial ELN-XLN configurations  with large neutrino-antineutrino asymmetry ($\langle D_0^z  \rangle\gtrsim 0.25$), we find that the angular crossing is erased by flavor conversion. However, for small neutrino-antineutrino asymmetries ($\langle D_0^z\rangle \lesssim 0.20$) and  more pronounced forward-peaked distributions,  new crossings can be created as the ``flipped" region of the angular distribution changes sign. For small neutrino-antineutrino asymmetry ($\langle D_0^z\rangle \lesssim 0.25$), flavor evolution is responsible for driving the system towards flavor equipartition.
Our results suggest that the final flavor configuration  strongly depends  on $\langle D_0^z\rangle $. Thus, it cannot be predicted relying on  arbitrary truncations of the multipole expansion of $D_v^z$. These findings {overcome} earlier approaches proposed in the literature, highlighting their limitations.

\begin{figure*}
\centering
\includegraphics[width=1\textwidth]{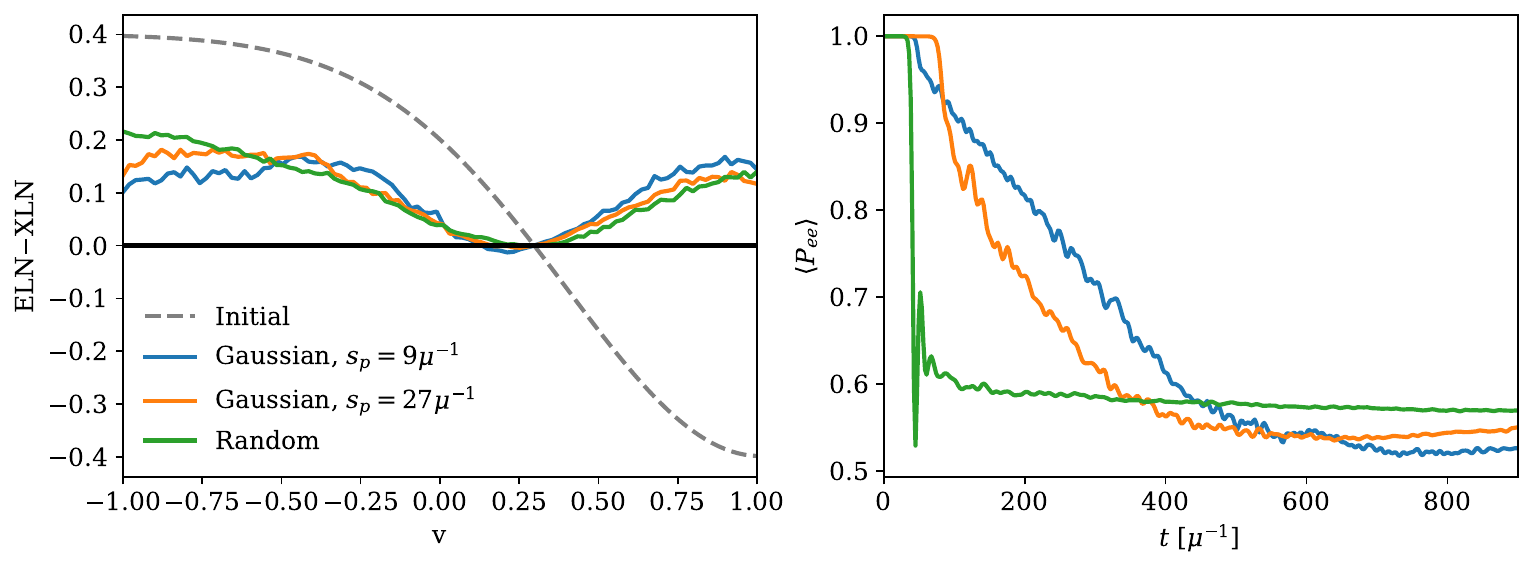}
\caption{\textit{Left:} Average ELN$-$XLN distribution at $t=900\mu^{-1}$ for the  flavor configuration C4 and three different initial seed perturbations. Independent of the perturbation, the $v>0$ part of the ELN-XLN distribution  changes sign as a result of fast flavor conversion. The initial perturbation mildly affects the shape of the distribution in the angular range with $v<0$. \textit{Right:} Space-averaged and angle-integrated $\nu_e$ survival probability as a function of time for the flavor configuration C4. The  initial perturbation affects  the onset of flavor conversion and the early evolution of the system, but the effect of the perturbation on the asymptotic $\langle P_{ee}\rangle$  is relatively small. }
\label{fig:5}
\end{figure*}

Although more work is required to investigate how our conclusions could be further modified by collisions with the background medium and vacuum effects, our results encouragingly suggest that simple analytical solutions can parametrize the steady-state flavor configuration. In addition, the phenomenology of neutrino self-interactions might be further modified in the core of supernovae or neutron-star merger remnants by the presence of large-scale gradients characterizing the quantities entering the equations of motion.
Despite this caveat, the investigation of neutrino transport in simplified setups  deepens our understanding of the kinetic equations and aids the search for cheap subgrid models of flavor instabilities, to be applied in large-scale source simulations.

\acknowledgments
{We thank Meng-Ru Wu and Masamichi Zaizen for useful discussions and comments on our manuscript.} This project has received support from  the European Union (ERC, ANET, Project No.~101087058) and the German Research Foundation (DFG) through the Collaborative Research Center ``Neutrinos and Dark Matter in Astro- and Particle Physics (NDM),'' Grant No.~SFB-1258-283604770. 
Views and opinions expressed are those of the authors only and do not necessarily reflect those of the European Union or the European Research Council. Neither the European Union nor the granting authority can be held responsible for them. The Tycho supercomputer hosted at the SCIENCE HPC Center at the University of Copenhagen was used to support our numerical simulations. 

\appendix

\section{Dependence of the steady-state flavor configuration on the choice of the initial perturbation}
\label{app:a}

{In this paper, flavor conversion is triggered by a  Gaussian perturbation described in  Eq.~\eqref{eq:perturbation}. This choice was motivated by its simple analytical form and its narrow Fourier spectrum. Other studies of neutrino quantum kinetics instead generate  random seeds of flavor coherence, see e.g.~Refs.~\cite{Wu:2021uvt, Zaizen:2022cik}. In this appendix, we investigate whether the steady-state flavor configuration depends on the choice of the initial perturbation. To this purpose, we consider the evolution of our initial configuration C4 (cf. Table~\ref{tab:1}) under three different initial conditions: two Gaussian perturbations of width  $s_p = 9 \mu^{-1}$ (used in Secs.~\ref{sec:flavor_ev} and \ref{sec:forecast}) and  $s_p=27\mu^{-1}$,  and a random perturbation $\rho_{v,xe}(0,r)=10^{-5}\epsilon(r)$, with  $\epsilon(r)\in [0,1]$.

Figure~\ref{fig:5} shows the angular distributions at { $t_f= 900\mu^{-1}$} (left panel) and the time evolution of the angle-integrated and space-averaged $\nu_e$ survival probability (right panel). The evolution of $\langle P_{ee}\rangle$ in the cases with Gaussian perturbations is qualitatively the same, whereas the random initial seed is responsible for a sharper transition towards the quasi-steady state. However, we find small differences in the final value of $\langle P_{ee}\rangle$. Hence,  Eq.~\eqref{eq:finalstate} remains a better prediction of the steady state  than \eqref{eq:equipartition} and Eqs.~\eqref{eq:no-crossing}. In particular, the angular range with $v>v_c$ is not driven to equipartition in any of our examples, as visible in the left panel of Fig.~\ref{fig:5}; rather, the  initial perturbation mainly affects the shape of the angular distribution in the range $v<0$. 

Reference~\cite{Wu:2021uvt} pointed out that the choice of the perturbation seed  can alter the evolution of  coherent flavor structures and found that point-like perturbations may result in larger flavor conversion than  random perturbations (cf.~their Fig.~6). 
Our results suggest that, although random perturbation seeds  lead to slightly higher average survival probabilities at late times, equipartition on one side of the ELN-XLN crossing is not a generic outcome of fast instabilities. Rather, it is a more common outcome
in ensembles that have a shallow initial ELN-XLN crossing. In general, we expect the effect of the initial perturbation on the asymptotic flavor configuration to be negligible, in agreement with findings obtained for  slow flavor conversion~\cite{Hannestad:2006nj,Duan:2007mv,Dasgupta:2010ae}.

\bibliography{advection}
\end{document}